\documentclass{PoS}

\title{Hard (State) Problems}

\ShortTitle{Hard (State) Problems}

\author{\speaker{John A. Tomsick}\\
        Space Science Laboratory, 7 Gauss Way, University of California, 
        Berkeley, CA 94720-7450, USA\\
        E-mail: \email{jtomsick@ssl.berkeley.edu}}

\abstract{For microquasars, the one time when these systems exhibit steady and
powerful jets is when they are in the hard state.  Thus, our understanding of this 
state is key to learning about the disk/jet connection.  Recent observational and 
theoretical results have led to questions about whether we really understand the 
physical properties of this state, and even our basic picture of this state is 
uncertain.  Here, I discuss some of the recent developments and possible problems 
with our understanding of this state.  Overall, it appears that the strongest 
challenge to the standard truncated disk picture is the detection of broad iron 
features in the X-ray spectra, and it seems that either there is a problem with the 
truncated disk picture or there is a problem with the relativistic reflection models 
used to explain the broad iron features.}

\FullConference{VII Microquasar Workshop: Microquasars and Beyond\\
		 September 1-5 2008\\
		 Foca, Izmir, Turkey}

\begin{document}

\section{Overview}

This work is devoted to a discussion of our current understanding of accreting
black hole systems when they are in the hard state.  In this work, I start
by describing the defining properties of the hard state and the questions that 
we would like to answer concerning the hard state.  Then, I discuss the
truncated disk picture, which most people consider to be the standard picture
for the hard state, and I present some of the key observational evidence in
favor of this picture.  Although the picture is successful in providing a
natural description for many observations, some recent observations appear
to be in conflict with the picture, and I describe some of these recent
observations.  One of the recent observations is presented our recent paper
entitled ``Broadband X-ray Spectra of GX~339--4 and the Geometry of 
Accreting Black Holes in the Hard State'' by Tomsick et al. (2008)
\cite{tomsick08}.  Finally, I will discuss some alternate pictures for the 
hard state as well as some on-going work to provide observational tests for 
theoretical hard state models.

\section{The Hard State}

For most black hole transients, the hard state is seen at the beginnings and
at the ends of outbursts.  It is often seen at relatively low mass accretion
rate, but mass accretion rate is not the only parameter that determines the
state of the system because hysteresis can occur where the transition from
the hard state at the beginning of an outburst typically occurs at a higher
luminosity level that the transition to the hard state at the end of the
outburst \cite{hb05}.  In addition, low-luminosity transients (e.g., 
XTE~J1118+480) can remain in the hard state for their entire outbursts, and
occasionally, a high-luminosity transient (e.g., V404~Cyg) can remain in the 
hard state for its entire outburst \cite{tomsick04_rossi}.  There are also
several persistent systems (Cyg~X-1, GRS~1758--258, 1E~1740.7--2942) spend 
most of their time in the hard state.

The hard state is defined by three main characteristics.  First, the energy
spectrum in the hard state is a power-law in the $\sim$5--100~keV bandpass
with a photon index in the range $\Gamma = 1.4$--2.0.  Although the 
($\nu F_{\nu}$) spectrum often peaks near 100~keV and is exponentially cutoff
above this energy \cite{grove98}, some observations suggest that the 
spectrum in this state can extend to higher energies without a cutoff
\cite{belloni06,caballerogarcia07}.  The second defining characteristic is a
high level of timing noise in the X-ray band.  The threshold value for the
hard state defined in \cite{mr06} is an rms noise level of 10\%, but many
systems show rms noise levels going up to $\sim$20--50\% \cite{pottschmidt03}.
The third defining characteristic is the presence of a steady and powerful
compact jet \cite{fender01}.  The compact jet is always seen in the radio 
band, and it has been resolved for GRS~1915+105 \cite{dhawan00} and Cyg~X-1
\cite{stirling01}.  There is strong evidence that the jet emission extends 
to the infrared and near-IR \cite{cf02,migliari07}, and it may extend to
even higher energies.

One of the main questions about the hard state that we would like to answer 
concerns the accretion geometry --- specifically, what is the inner radius
($R_{\rm in}$) of the optically thick disk.  This is important for understanding
the relationship between the optically thick disk and the corona (i.e., the
site of the hard X-ray production) as well as constraining the disk/jet
connection.  One can imagine that the geometry of the optically thick disk
may be very important for creating the conditions necessary for the 
production of detectable compact jets.  For example, the jet may rely on
matter coming from the optically thick disk, suggesting that a small
$R_{\rm in}$ may be required to feed the jet.  On the other hand, if there
is too much soft X-ray emission incident on the jet, then this can cause
significant Compton cooling of the jet electrons, and this suggests that
if the disk remains relatively hot (and bright) that a relatively large
$R_{\rm in}$ may be required to avoid cooling the jet.  Either way, it is 
clear that one expects that the interaction between the optically thick 
disk and the jet to be a significant one, making it important to determine 
the accretion geometry.

\section{The Truncated Disk Picture}

The idea that the hard X-ray emission in the hard state comes from an
optically thin region close to the black hole while the optically thick
disk recedes was conceived early-on in the study of accretion disks around
black holes \cite{sle76}.  Although the Shapiro, Lightman, \& Eardley
(1976) model proved to be unstable, this sort of accretion geometry is 
viable if advection is considered as in Advection-Dominated Accretion 
Flow (ADAF) models \cite{ny94}.  It was shown that the ADAF model could
generally explain the energy spectra seen for the different black hole
states by allowing for changes in mass accretion rate and $R_{\rm in}$
\cite{emn97}.  Importantly, this model envisions that $R_{\rm in}$ makes
its largest changes as it passes through the intermediate state on its
way to the hard state.  To be viable, this model requires that 
$R_{\rm in}$ becomes very large ($\sim$100--1000 gravitational radii, 
$R_{\rm g}$) when systems are in the hard state and in quiescence.
Although the Esin et al. (1997) work does not consider the presence of
jets, a recent review article by Narayan \& McClintock (2008) \cite{nm08} 
shows the basic ADAF version of the truncated disk picture.  (I note that 
while this is a good representation of the basic picture, one aspect of 
the picture that does not match the observations is the indication of a 
compact jet in the intermediate state.  In fact, it has been shown that, 
during outburst decay, the jet does not turn on while the spectrum is 
evolving in the intermediate state.  It only turns on when the source 
reaches its hardest level \cite{kalemci05}.)

In the following subsections we list several observations that are 
well explained within the truncated disk picture.

\subsection{Changes to the Energy Spectrum}

As black hole systems pass from the soft (or ``thermal-dominant'' or TD) 
state to the hard state, the primary emission component changes from being 
a thermal component from the optically thick disk to the hard power-law 
component discussed above.  In almost all systems, when the thermal 
component is modeled by a disk-blackbody, the inner disk temperature, 
$kT_{\rm in}$, drops from values near 1~keV in the TD state to $<$0.3~keV
in the hard state \cite{kalemci04}.  After the drop in temperature, 
the thermal component is often not detected, especially if the X-ray
coverage only includes $>$3~keV (e.g., the {\em Rossi X-ray Timing 
Explorer (RXTE)}).

In most cases, it is not clear whether the drop in $kT_{\rm in}$ is due to
a change in $R_{\rm in}$ or a change in mass accretion rate because for most
black hole systems, the extinction is too high to obtain a good measurement
of the shape of the thermal component since most of the emission is in the
soft X-ray and UV bands.  However, good measurements were obtained for
one system, XTE~J1118+480, because it is relatively nearby, $\sim$1.8~kpc, 
and it is at a high Galactic latitude of $b = +62^{\circ}$, so its extinction
is very low.  At 0.1\% of the Eddington luminosity ($L_{\rm Edd}$), the optical, 
UV, and X-ray spectrum was measured, and the optical an UV emission was 
well-described by a disk-blackbody component with $kT_{\rm in} = 0.02$~keV.  
Although different studies which modeled this component found different 
values for $R_{\rm in}$, the values found were between 110 and 700 $R_{\rm g}$ 
\cite{esin01,chaty03,ycn05}, which are consistent with expectations for 
the truncated disk model.

A second spectral feature that is consistent with the truncated disk
model is related to the hard power-law tail.  If this emission is due
to inverse Compontonization by a thermal distribution of electrons, then
the corona where the hard X-ray emission is produced must be photon-starved
in order to maintain a temperature as high as 100~keV \cite{done07}.
While this is consistent with a truncated disk, and a truncated disk seems
to be the most natural explanation, other possibilities include a drop in 
the radiative efficiency in the disk or other geometrical changes that 
might somehow isolate the corona from soft X-ray emission produced by
the optically thick disk.

\subsection{Drop in Characteristic Frequencies}

The evolution of the timing properties during state transitions also seem
to find a natural explanation within the truncated disk picture.  Both 
Quasi-Periodic Oscillations (QPOs) and other characteristic frequencies
detected from black hole systems, such as the break frequency in the power 
spectrum, exhibit a decrease as the mass accretion rate drops during
the transition to the hard state and in the hard state.  Furthermore, 
several groups have pointed out that there is a strong correlation
between the QPO frequency and the power-law index, $\Gamma$
\cite{gcr99,kalemci_thesis,vignarca03,tomsick04_rossi,st07}, connecting
the timing changes to spectral changes that are well-explained by a
truncated disk.  The example shown in \cite{tomsick04_rossi} shows that 
the typical behavior is for the QPO frequency to change from $\sim$10~Hz
to $\sim$0.2~Hz as $\Gamma$ changes from 2.4 to 1.6.  If this frequency
scales with the dynamical time scale at $R_{\rm in}$, then such a change
in frequency would imply a change in $R_{\rm in}$ by a factor of 14, so
that if $R_{\rm in}$ was at 6 $R_{g}$ in the TD state, then it would be
at 84 $R_{g}$ by the time the system reached the hard state.  Although
the QPO is usually not detected in fainter hard states, there is 
evidence that characteristic frequencies continue to drop in the hard 
state as flux decreases \cite{tkk04}.

\subsection{Drop in the Strength of the Reflection Component}

Another spectral feature that is strongly dependent on the geometry of
the system is the reflection component, whose strength depends on the
solid angle subtended by the optically thick disk as seen from the hard
X-ray source, $\Omega/2\pi$.  The Compton reflection component consists
of a bump in the emission bewtween $\sim$20 and 100~keV \cite{lw88}
as well as absorption edges and fluorescent emission lines, most prominently
from iron.

An important indication that the accretion geometry changes as black hole
systems enter the hard state is that $\Omega/2\pi$ shows a dramatic drop.
It was shown that several black hole systems show very similar changes, 
specifically, that $\Omega/2\pi$ changes from $\sim$1 to $\sim$0.3 as 
$\Gamma$ changes from 2.2 to 1.5 \cite{zls99,zdziarski03}.  It was also
shown that this evolution can be well explained by the truncated disk
picture along with a geometry where there is some overlap between the
corona and the optically thick disk \cite{zdziarski03}.  This explanation
is not necessarily a unique explanation for the evolution, but the good
agreement between this model and the data indicate that it is a possible
explanation.

\section{Recent Observations that may Conflict with the Truncated Disk Picture}

While the observations described in the preceeding section can be seen
as supporting the truncated disk picture, some recent observations 
present challenges to this picture.  In part, improved X-ray instrumentation
over the past several years has provided this new information, including the
better soft X-ray sensitivity and/or better energy resolution of 
{\em XMM-Newton}, the {\em Chandra X-ray Observatory}, {\em Swift}, and
{\em Suzaku} when compared with previous satellites.  Here, the two
challenges we describe and discuss are measurements of the thermal disk
component in the hard state and the broad iron features in the reflection
component.

\subsection{Thermal Disk Components}

Although it has long been known that sources in relatively bright hard states
can exhibit thermal disk components with temperatures of $kT_{\rm in}\sim 0.2$ keV,
recent observations have shown that this component is present in more sources
and at lower luminosities than seen previously.  For Cyg~X-1, this component
has been detected by satellites from {\em ASCA} \cite{ebisawa96} to 
{\em BeppoSAX} \cite{ddz01} to {\em Suzaku} \cite{makishima08} when the
source has been at a luminosity of $\sim$2\% $L_{\rm Edd}$.  It has also been
detected at around this luminosity for GX~339--4 and GRO~J1655--40, but there
have been reports that it persists at a level of $kT_{\rm in}\sim 0.2$~keV to 
much lower luminosities, $\sim$0.1\% $L_{\rm Edd}$, for Swift~J1753.5--0127 and 
XTE~J1817--330 \cite{miller06b,rykoff07}.

From modeling these components, it has been suggested that they are consistent
with little or no change in $R_{\rm in}$ relative to the location of the inner
radius in the TD state \cite{miller06b,rykoff07}.  The finding in the case
of XTE~J1817--330 is that the evolution of the properties of the thermal
component during $\sim$20 {\em Swift} observations is consistent with 
an $L\propto T_{\rm in}^{4}$ relationship, implying that any changes in the
thermal disk component is related to a change in mass accretion rate rather
than a change in $R_{\rm in}$.  However, obtaining physically meaningful
parameters by fitting thermal disk components is challenging, and there
are many uncertainties that can introduce systematic errors such as the
assumptions about the inner boundary condition, the viscosity prescription, 
electron scattering that can cause a difference between the effective
temperature and the color temperature, and changes to the temperature 
profile in the disk due to irradiation.  In fact, when the same {\em Swift}
observations that were used to conclude that the inner radius does not
change, were re-analyzed accounting for a different inner boundary
condition and irradiation of the disk, it was found that the inner radius
could change by as much as a factor of 10 \cite{gdp08}.

In our recent work on {\em Swift} and {\em RXTE} observations of 
GX~339--4 in the hard state, we also detect a thermal disk component at
$kT_{\rm in} = 0.193\pm 0.012$~keV at a luminosity of 2.3\% $L_{\rm Edd}$.  
The normalization of the disk-blackbody component is related to the
inner radius according to $N_{\rm DBB}\propto R_{\rm in}^{2}$, and we 
measure a value of $N_{\rm DBB} = (2.9^{+2.7}_{-1.4})\times 10^{4}$
\cite{tomsick08}.  GX~339--4 was also observed at a higher luminosity, 
5.6\% $L_{\rm Edd}$, and a higher thermal disk temperature was measured, 
$kT_{\rm in} = 0.39\pm 0.04$~keV.  However, the value of $N_{\rm DBB}$
of $700\pm 200$ is much lower than seen in our observation of GX~339--4
\cite{miller06a}, which could be taken as evidence that $R_{\rm in}$ 
increases between the two observations.  While this could be the case, 
this conclusion is not consistent with the iron line measurements 
reported in \cite{tomsick08} and \cite{miller06a} and discussed below, 
and it is not clear how large the systematic error is on the measurement
of $R_{\rm in}$ using the thermal disk component.

Overall, my conclusion concerning using thermal disk components to 
constrain the evolution of $R_{\rm in}$ is that it is difficult to be
sure that the systematic errors are well-understood.  However, it is
interesting that sources like Swift~J1753.5--0127 and XTE~J1817--330
can show disk temperatures of $\sim$0.2~keV while XTE~J1118+480 
has a disk temperature that is an order of magnitude lower, 0.02~keV
at the same luminosity level, $\sim$0.1\% $L_{\rm Edd}$.  Even if this
does not give a clean measurement of $R_{\rm in}$, efforts to improve
our understanding of this component are still worthwhile.

\subsection{Broad Iron Features in the Hard State}

The second challenge to the truncated disk model is the detection of
broad iron features in the hard state.  As discussed above, these 
features are part of the Compton reflection component, and the leading
interpretation for the width of the features is that they are 
relativistically smeared by motion of accretion disk material around
the black hole, causing Doppler shifts, as well as by redshifts caused
by the black hole's gravitational field.  Although such features have
been seen in several Galactic black hole systems when they are at
high luminosities, the hard state has not been as well-studied with
the higher energy resolution available with the current generation of
X-ray satellites.  However, there are now a few hard state iron line 
studies that use higher resolution instrumentation along with spectral
models that include the relativistic effects.  The main source that 
has been studied in this way is GX~339--4, and here we describe results
that have been obtained for this source with {\em XMM-Newton}, {\em Swift}, 
and {\em RXTE} over the past few years.  We note that in deriving the
Eddington-scaled luminosities, we assumed a distance of 8~kpc for
GX~339--4 and a black hole mass of 5.8 solar masses.

\begin{figure}
\centerline{\includegraphics[width=1.0\textwidth]{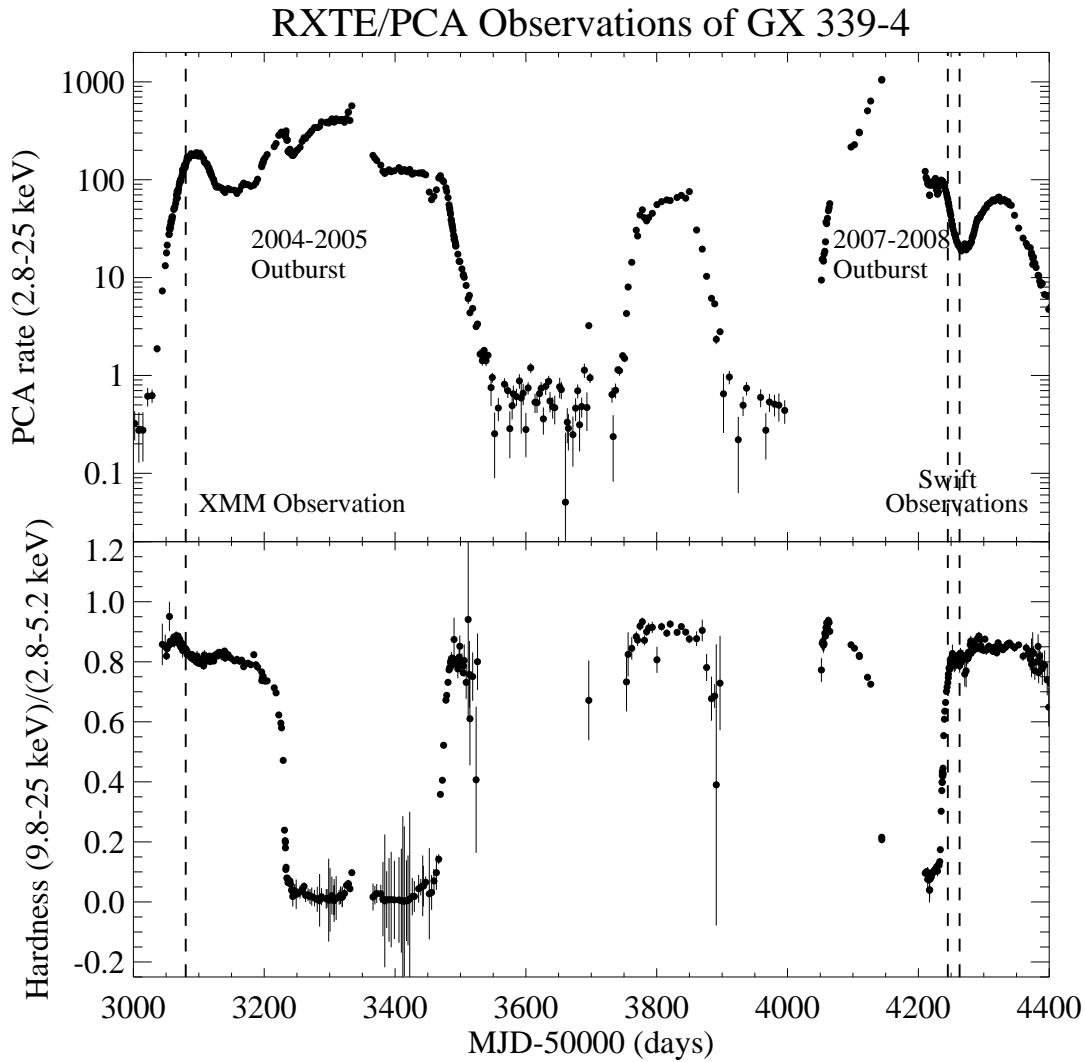}}
\caption{GX 339--4 light curve and hardness ratio from pointed observations
with the {\em Rossi X-ray Timing Explorer}'s Proportional Counter Array (PCA)
instrument.  Three separate outbursts were detected between 2004 and 2008.  
The vertical dashed lines mark an {\em XMM-Newton} observation in 2004 and
two {\em Swift} observations in 2007.}
\label{fig:pca339}
\end{figure}

Figure~\ref{fig:pca339} shows the {\em RXTE} Proportional Counter Array
(PCA) light curve and hardness ratio for GX~339--4 from 2004--2008.  The
system normally has a high level of activity, but the amount of activity
over this 4 year period is exceptionally high with 3 separate outbursts.
The first study of iron features was carried out using {\em XMM-Newton}
and {\em RXTE} data from 2004.  The observation occurred at the start
of the 2004--2005 outburst at a relatively high luminosity, 5.6\%
$L_{\rm Edd}$, and the hardness ratio shows that the system was in the
hard state.  The results of the {\em XMM-Newton} plus {\em RXTE} spectral
analysis are reported in \cite{miller06a}.  The spectra show evidence
for broad iron features, and they are fitted with relativistic smearing
(the ``kdblur'' model in XSPEC) and two different reflection models
(``pexriv'' and the ``constant density ionized disk'' or ``CDID'' model).
The result is that the smearing of the iron features is strong enough 
to require that the disk extends very close to the black hole.  With 
the first reflection model, $R_{\rm in}$ is constrained to be 
$4.0\pm 0.5$ $R_{\rm g}$ while $5.0\pm 0.5$ $R_{\rm g}$ is obtained with
the CDID model.  

\begin{figure}
\centerline{\includegraphics[width=1.0\textwidth]{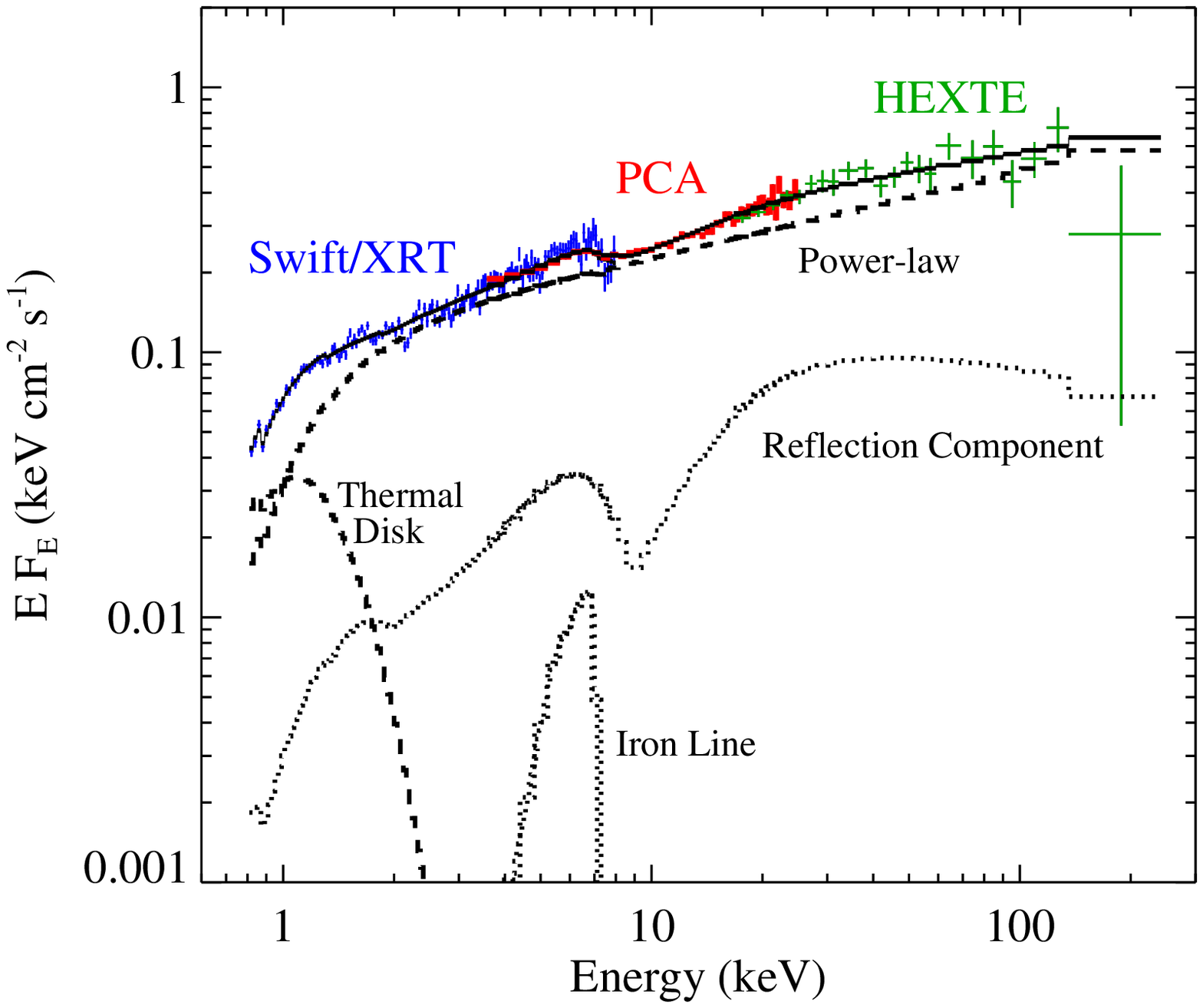}}
\caption{The {\em Swift} and {\em RXTE} energy spectrum of GX~339--4 taken 
in 2007 May when the source was in the hard state at a luminosity of 
2.3\% $L_{\rm Edd}$.  The model components are shown, and we focus on the
thermal disk component and the broad reflection features in this work.
The figure has been adapted from Tomsick et al. (2008) \cite{tomsick08}.}
\label{fig:spectrum}
\end{figure}

A separate study of GX~339--4 was carried out by our group.  We used the
{\em Swift} and {\em RXTE} observations marked in Figure~\ref{fig:pca339}, 
and it is notable that these observations were carried out at the end
of the 2007--2008 outburst (in contrast to the {\em XMM-Newton} 
observation described above), and the {\em Swift} observations occurred
at lower luminosity, 2.3\% and 0.8\% $L_{\rm Edd}$, than the earlier
observation.  Even so, the {\em Swift} and {\em RXTE} spectra also show
strong evidence for broad iron features \cite{tomsick08}.  Our result
using the pexriv reflection model smeared by kdblur is that $R_{\rm in}$
is constrained to be $3.6^{+1.4}_{-1.0}$ $R_{\rm g}$ at 2.3\% $L_{\rm Edd}$
and $<$10 $R_{\rm g}$ at 0.8\% $L_{\rm Edd}$.  The spectrum at the 
higher luminosity level is shown in Figure~\ref{fig:spectrum}.

Although GX~339--4 is the only system with published studies that have used
the most recent instrumentation and relativistic reflection models, broad
iron features have also been reported for Cyg~X-1 \cite{makishima08} and
for XTE~J1650--500 \cite{dg06}.  Thus, the features are not simply 
restricted to GX~339--4.  For GX~339--4, the small inner radii derived
are in contradiction to what is expected for the truncated disk model, 
and this implies that, at least down to 2.3\% $L_{\rm Edd}$, either the
truncated disk picture is wrong or we are mis-interpreting the broad
reflection features.

\section{Some Alternate Pictures}

Other hard state accretion geometries have been suggested, and it is
worthwhile to consider whether they may provide a better explaination 
for the observations than the truncated disk model.  Here, I will 
consider a compact jet model and a picture where an inner cool disk 
remains in the hard state.

\subsection{Compact Jet Model}

First, it should be noted that the compact jet model is not necessarily
incompatible with the truncated disk picture since it is possible that the
compact jet is fed by an ADAF.  However, a version of the compact jet 
model that may be able to explain the broad iron features would have the 
optically thick disk extending within several $R_{g}$ of the black hole 
while the jet is also launched and collimated in this region.  In addition, 
if a significant fraction of the X-ray emission came from the base of the 
jet as envisioned by Markoff, Nowak, \& Wilms (2005) \cite{mnw05}, then 
this would give a reflection component from the inner part of the disk 
that could explain the broad iron features.

In fact, the Markoff et al. \cite{mnw05} compact jet model has now been 
applied to hard state Spectral Energy Distributions for a few black hole 
systems.  One parameter in the model is the ``nozzle radius,'' which is 
the radius of the base of the jet.  In the spectral model, this parameter 
is related to the location of the break from optically thick to optically
thin synchrotron emission, and it has been relatively well-constrained
by some SEDs.  When the SEDs for Cyg~X-1, GX~339--4, and GRO~J1655--40
have been fitted, values for this parameter have ranged from 3.5 to
9.6~$R_{g}$ \cite{mnw05,migliari07}, which is similar to the values 
of $R_{\rm in}$ that are derived from the relativistic reflection models.  

However, even if such a picture appears to be consistent with the
broadened iron features, it is still necessary to ask whether it can 
explain the overall strength of the reflection component, the shape
of the hard X-ray spectrum, and the changes in characteristic frequencies.
Although some have argued that such a picture would lead to a much
higher value for $\Omega/2\pi$, Markoff \& Nowak (2004) \cite{mn04} have 
shown that the model is consistent with a drop in this parameter as 
sources make transitions to the hard state because the X-ray emission in 
the jet will be relativistically beamed away from the optically thick 
disk, explaining the drop in $\Omega/2\pi$.  Concerning the shape of 
the X-ray spectrum, it is important to determine if black hole spectra 
always have an cutoff near 100 keV or not.  However, we do know that the 
spectra sometimes have exponential cutoffs near this energy \cite{grove98}.
The compact jet model does predict cutoffs in the spectra, but it 
would be useful to determine if the model can specifically predict
some of the hard X-ray spectra where the cutoff shape has been very
well-measured.  Concerning the characteristic frequencies, it is 
currently an open question whether QPOs would be expected for an
X-ray emitting compact jet.

\subsection{Inner Cool Disk Picture}

Another possibility that has been explored in some detail recently is
the idea that an inner optically thick disk can remain in the hard
state.  As the mass accretion rate drops during the transition to
the hard state, the truncated disk picture envisions that material 
from the optically thick disk will evaporate into the corona starting 
from the inner edge of the disk.  However, calculations by \cite{liu02}
that account for viscous generation of heat in the corona and
conduction of that heat into the optically thick disk show that 
the maximum evaporation rate should actually be at $\sim$50 to
a few hundred gravitational radii (depending mostly on viscosity
in the corona, i.e., the $\alpha$ parameter).  Since the inner disk
has the strongest level of soft photon emission, when Compton 
cooling of the corona is also accounted for, the effect becomes 
even more pronounced \cite{liu02}.  Thus, not only is it possible
for a inner cool disk to remain, but it is actually what calculations
predict should happen.

Recently, \cite{liu07} and \cite{taam08} have applied the inner cool
disk model to observations by determining whether the model can 
reproduce the thermal disk components seen in GX~339--4 and 
Swift~J1753.5--0127.  In fact, it was found that for viscosity
parameters in the range $\alpha\sim 0.1$--0.4, the thermal disk
components could be reproduced.  For the 2007 observations of
GX~339--4 with a thermal disk temperature of $T_{\rm eff} = 0.203$~keV
and a luminosity of 2.3\% $L_{\rm Edd}$ \cite{tomsick08}, it was
shown that the observations could be reproduced with 
$\alpha = 0.31$, an Eddington-scaled mass accretion rate of 6.87\%, 
and an inner optically thick cool disk extending from the 
marginally stable orbit (i.e., 1--6 $R_{\rm g}$, depending on the
black hole's spin) to 142 $R_{\rm g}$.  

While it has been shown that the model can explain the thermal
disk component in the hard state, it has not yet been determined
whether it could explain the broadened iron lines.  Since the
optically thick disk extends to the marginally stable orbit in 
the model, reflection from that region should certainly produce
the broad iron features that are observed.  However, it is not
clear if the geometry of the hard X-ray source is consistent 
with most of the hard X-rays illuminating the inner part of the
disk.  This would be a very useful theoretical study to carry
out.

Finally, it is a prediction of the inner cool disk model that 
when the mass accretion rate becomes low enough the entire inner
disk will eventually evaporate.  The calculations of \cite{liu07}
and \cite{taam08} suggests that the complete evaporation of the
inner part of the disk should occur at a luminosity near 0.1\%
$L_{\rm Edd}$.

\section{Summary and Future Work}

{\em The most important point is that it really seems like the 
observations are telling us that there is a problem either with the 
truncated disk picture or with our interpretation of the broad iron 
features for black holes in the hard state.}  While thus far, the 
observations only indicate a problem down to a luminosity of 
$\sim$1\% $L_{\rm Edd}$, there are strong implications for our
entire understanding of the differences between the most 
fundamental black hole spectral states: Soft (or TD) and hard.
Although the thermal disk component may also indicate a problem
down to 0.1\%, the evidence that this component requires a 
non-truncated disk is weaker than the broad iron features.

Thus, it is very important to determine if the relativistic
interpretation for the broad iron features is the only viable
one.  It has been suggested that the red wing of the iron line
could be explained by Compton down-scattering of the line
emission \cite{lt07}; however, it is still not clear if this
model can simultaneously explain the other properties of the 
hard state.  Also, it has been mentioned that it is not clear
if the relativistic models are adequate for the case of stellar
mass black holes.  Although they are often applied to the 
stellar mass case, the models mentioned above (pexriv and CDID)
were originally developed for Active Galactic Nuclei (AGN).  
Some inaccuracies may be present when using pexriv for hotter
accretion disks because of the way it deals with higher ionizations.
While the CDID model deals with hot disks (thermal motions and
higher ionization states) in a much improved manner, it is 
still true that the matter densities that are assumed are
appropriate for AGN.  Developing a relativistic reflection
model specifically for stellar mass black holes would be a
positive step.

Also, more can be done on the observational side.  We currently
have programs to try to extend the hard state studies to lower
luminosities with {\em XMM-Newton}, {\em Suzaku}, and {\em Swift}.
In particular, we recently (2008 September) carried out a 100~ks 
{\em Suzaku} observation of GX~339--4 at a significantly lower 
luminosity than the \cite{tomsick08} {\em Swift} and {\em RXTE} 
study.  It will be very interesting to see if we find any change
in the broad iron features. 

\vspace{5mm}
\emph{Acknowledgements}
I would especially like to thank Emrah Kalemci for all his efforts 
in hosting this workshop.  I would also like to thank the Local and
Scientific Organizing Committees for helping to make the meeting a
great success.  I acknowledge many contributions from my collaborators, 
including Stephane Corbel, Emrah Kalemci, Phil Kaaret, Sera Markoff, 
Simone Migliari, Rob Fender, Charles Bailyn, and Michelle Buxton.  
I also acknowledge useful discussions with participants of the 
workshop.  I acknowledge partial support from NASA {\em RXTE} Guest 
Observer grants NNG06GA81G and NNX06AG83G and from NASA {\em Swift} 
Guest Observer grant NNX08AQ60G.



\end{document}